\documentclass{aa}
\usepackage{graphics}
\begin{document} 
\thesaurus{
              (08.06.2  stars:formation
	       08.16.5  stars:pre-main sequence
	       09.08.2  Herbig-Haro objects 
               09.10.1  ISM: jets and outflows	
               13.09.6  Infrared:stars
               02.01.2  Accretion, accretion disks
               )} 

\title{Shocked Molecular Hydrogen from RNO\,91} 

\author{ M. S. Nanda Kumar\inst{1}, B. G. Anandarao \inst{1} \& C.J. Davis
\inst{2}}
\institute{Physical Research Laboratory, Navrangpura, Ahmedabad, India -
380009   \\ e-mail:nanda@prl.ernet.in,anand@prl.ernet.in 
\and
 Joint Astronomy Center, 660 N. A'oh\={o}k\={u} Place, 
University Park,Hilo, HI 96720, USA \\e-mail:cdavis@jach.hawaii.edu}

\offprints{M. S. Nanda Kumar}
\date{6october1998}
\authorrunning{Nanda Kumar et al.}
\titlerunning{Shocked H$_{2}$ emission from RNO\,91}
\maketitle

\begin{abstract} 

We report the detection of the H$_{2}$ $\it{v}$=1-0 S(1) line at
2.122$\mu$m, from RNO\,91 in the L43 dark cloud, which is known to be a T
Tauri star surrounded by a 1700AU disk structure (containing ices)
and a weak outflow. The non-detection of the H$_{2}$ $\it{v}$=2-1 S(1)
line at 2.247$\mu $m suggests shock excitation rather than fluorescence.
The emission is extended spatially up to 9$\arcsec$ in the north-south
direction. The line intensity peak (FWHM $\sim$ 3$\arcsec$) corresponds to
the star RNO\,91 which is embedded in a cocoon of gas and dust. The
observed H$_{2}$ emission from this cocoon may be attributed to embedded
Herbig-Haro like knots. The H$_{2}$ line flux in the central $2\arcsec
\times 3\arcsec$ is estimated to be $7 \times 10^{-14}$ ergs
sec$^{-1}$cm$^{-2}$, which indicates a mass flow rate of $4\times 10^{-8}$
M$_{\odot}$ yr$^{-1}$. Furthermore, narrow band image taken through H$_{2}$
1-0 S(1) filter is presented, which reveal a tilted disk and bipolar
outflow structure that agrees with earlier observations and models. We
show that this disk/outflow system is a unique case.

\end{abstract}

\section{Introduction} 

The near-IR molecular hydrogen emission lines are recognised as being
important tools in studies of star formation (Shull \& Beckwith
\cite{shul}). The excitation of these lines involves mainly two competing
processes: (i) shock heating and (ii) UV fluorescence (Burton
\cite{bur92}). It is possible, however, to distinguish between these two
processes by measuring the ratios of intensities of lines arising from two
different vibrational levels (Sternberg \& Dalgarno \cite{ster}, Hora \&
Latter \cite{hor}). The origin of shocked molecular hydrogen emission from
the spatially unresolved region close to the star can originate from
outflows or from accretion shocks in a disk, as in the case of Infrared
Companions (IRC's) (Herbst  et al. \cite{her95}, Koresko et al.
\cite{kor}).  RNO\,91($\alpha = 16^{\rm h} 34^{\rm m} 29.3^{s}, \delta =
-15^{\circ} 47\arcmin 01\arcsec$)is one of the only two known PMS stars in
the L43 dark cloud in Ophiuchus. It was classified as a M0.5 type T Tauri
star by Leverault (\cite{lev}), based on an optical spectrum that showed
strong H$\alpha$ emission. An outflow driven by this star was identified
at millimeter wavelengths and shown to have spatially separated
red-shifted and blue-shifted lobes (Leverault \cite{lev}, Myers et al.
\cite{myr}, Bence et al. \cite{ben}). However, optical images and spectra
obtained by Schild et al. (\cite{sch})  showed that the outflow does not
have any emission indicative of shocked material around RNO\,91. Optical
and infrared photometry (U-band to L-band) was obtained by Myers et
al.(\cite{myr}). Heyer et al. (\cite{heyr}) obtained the JHK photometry
and H-band polarimetry of this object which showed a disk type structure
for the first time. Weintraub et al. (\cite{wein}, hereafter, W94)
demonstrated by their K-band polarimetric image and 3-5 $\mu$m spectra
that RNO\,91 is surrounded by a disk-like structure of radius 1700 AU
comprising frozen H$_{2}$O, CO and possibly XCN. They also showed that the
polarization center does not coincide with the intensity peak identified
as RNO\,91. Our speculation that  this result indicates  an IRC to RNO\,91
was quickly laid to rest by the Shift and Add  imaging by Aspin et al.
(\cite{asp}) (hereafter A97), which does not show any secondary source
within a 3.7$\arcsec$ square region. In this paper we present
near-infrared spectra and narrow band images of the object in the K band
region. We also report the detection of the H$_{2} $ $\it{v}$= 1-0 S(1)
line at the source and discuss its implications.  

\section{Observations and Data Reduction}

Near infrared spectroscopic observations were made on March 25, 1998 at
Gurushikhar 1.2 m Infrared Telescope (GIRT), Mt. Abu, India. A Near
Infrared Camera / Grating Spectrometer based on a HgCdTe $256\times256$
focal plane array was used to obtain the observations. The grating
spectrometer was used in a configuration that yielded a resolving power of
$\lambda / \delta \lambda$ = 1000 with a 1$\arcsec$/pixel plate scale. 
The slit was two pixels wide and oriented along the N-S axis. The
atmospheric seeing and image motions were below $2\arcsec$ during the
observations, which was measured from imaging data obtained just
before the spectroscopic observations. Data acquisition and reduction
were done using standard procedures. We used the RNO\,91 K band
photometric fluxes (0.48\,Jy) given by Myers et al. (\cite{myr}) for flux
calibrating the spectrum.

Narrow band ($ \Delta\lambda =0.02 \mu$m) images through
H$_{2}$ ($\lambda=2.122\mu$m),  Br$\gamma$($\lambda=2.165 \mu$m) and
continuum ($\lambda=2.104 \mu$m) filters were obtained by the United
Kingdom Infrared Telescope (UKIRT) Service Observing Program on September
8, 1998  using the facility near-IR imager IRCAM3. IRCAM3 employs a
$256\times256$ InSb array; the optics used gives a pixel scale of
0.280\arcsec. There was a small defocusing problem that occurred during
these observations due to variable seeing, resulting in image elongation
in the N-E, S-W direction. These uncertainties were estimated to be about
1.2$\arcsec$. Continuum subtraction was not carried out because of the
focusing problems in these images. It should also be noted that there is
some ghosting that had occurred in these images. These are identified as
ghost images since they occur at exactly the same position with respect to
the main source both in object and standard star frames.

\begin{figure}
\resizebox{\hsize}{!}{\includegraphics{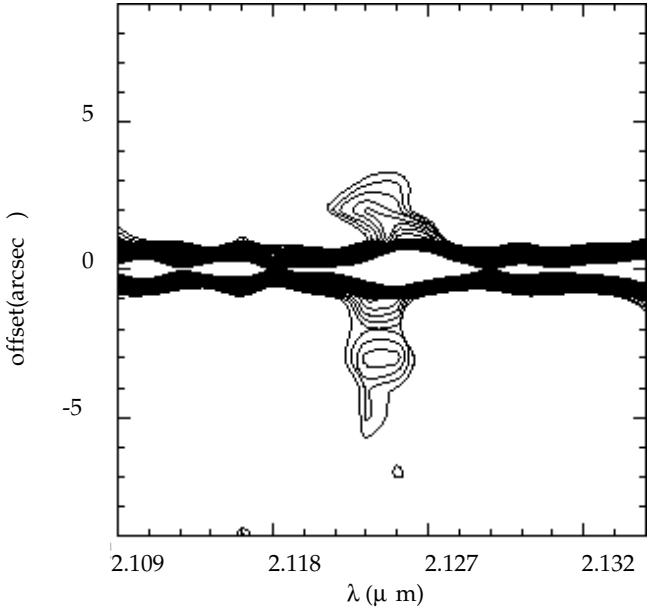}}
\caption[]{ Contours of {H$_{2}$ 2.122$\mu$m }line from  RNO\,91  spectrum
showing the extended emission. Notice the different lengths of emission in
the N-S from the center that indicates the  tilt of the outflow axis.
Continuum is {\it partly subtracted} using an adjacent strip of the same
spectrum, which has caused the artifact of width variability.}
\label{mag1}
\end{figure}

\section{Results} 

The contour map of the spectrum shown in figure 1 displays the extended
H$_{2}$ emission along the N-S direction (the slit axis) corresponding to
the outflow axis.  The intensity peaks at -3$\arcsec$ and +2$\arcsec$
represent H$_{2}$ knots in the outflow. From gaussian fits to the
continuum emission profile at different positions along the dispersion
axis, we find that the FWHM of the continuum strip (measured N-S) is about
3.0$\arcsec$, although the seeing on the night of observation was between
1.5-2.0$\arcsec$.  The extra width in the stellar continuum strip could be
attributed to scattered light from the cocoon surrounding the star.  We
have therefore extracted the source spectrum by integrating the emission
along three and then five rows, representing an on-source area of
$2\arcsec \times 3\arcsec$ and $2\arcsec \times 5\arcsec$.  The relative
intensities of the photospheric NaI and CaI features remained the same in
these two extractions.  However, in the $3\arcsec$-wide extraction, the
2.122$\mu$m line intensity was considerably smaller relative to the
photospheric features. This result confirms that the broadening of the
continuum strip is indeed due to scattered star light.

\begin{figure}
\resizebox{\hsize}{!}{\includegraphics{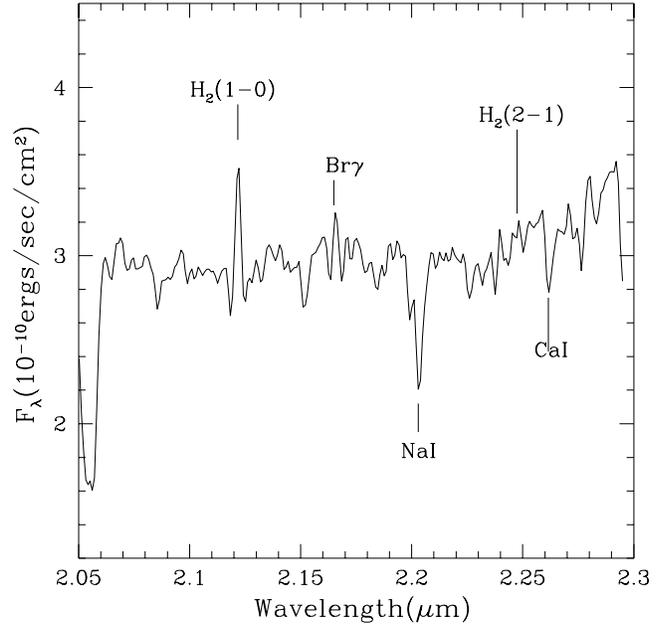}}
\caption[]{Spectrum of RNO\,91 integrated over 9 rows representing an
area on-source of 2$\arcsec \times$ 9$\arcsec$.}
\label{mag2}
\end{figure}

Figure 2 shows the spectrum of RNO\,91 in the wavelength region $2.05\mu
$m to $2.3\mu$m. The spectrum is obtained by integrating nine rows,
covering 9$\arcsec$ along the N-S slit axis centered around the star. The
spectrum displays prominently the H$_{2}$ $\it{v}$=1-0 S(1)line at $2.122\mu$m
and the photospheric NaI and CaI absorption features at $2.20\mu$m and
$2.26\mu$m respectively. The 2-1 S(1) line at 2.247$\mu$m is below the
noise level. The spectrum also displays the Br$\gamma$ emission line at
2.167$\mu$m. These features are marked in the figure. Most of the other
features seen in absorption are telluric in nature (Chelli et al.
\cite{chel}).

The  excitation mechanism for H$_{2}$ emission can be inferred from the
2-1 S(1)/1-0 S(1) flux ratio (e.g. Luhman et al. \cite{luhm}). An
estimated upper limit to the 2-1 S(1) line flux yields this ratio to be
0.16 or less, pointing strongly towards shock excitation. However,
flourescent excitation  in a high density regime may also produce a
``shock-like'' 2-1 S(1)/1-0 S(1) ratio, because of thermalisation of
low-energy vibrational levels (Sternberg \& Dalgarno \cite{ster}, Burton
et al. \cite{bur90}). In a high density region one would still expect to
see emission from the $\it{v}$=3 level at 10\% of the 1-0 S(1) line
(rather than 1\%, as is expected in a shock; see e.g. Luhman et al.
\cite{luhm}). However, in our spectrum the 3-2 S(3) line is unresolved from
NaI absorption, and the 3-2 S(1) line is outside the wavelength
range. 

We resolve this issue by estimating the two most crucial parameters that
decide the efficiency of the UV fluorescence, namely, the gas number
density and the UV flux scaling parameter $\chi$ (see Sternberg and
Dalgarno \cite{ster}). We estimate an upper limit for the gas number
density to be $6\times 10^{5} $ cm$^{-3}$, considering an A$_{v} \sim$ 9
(Myers et al. \cite{myr}), an outer disk radius of 1700AU (W94), and an
inner disk radius of 0.01AU corresponding to the dust evaporating radius
for an M0 star. Note that this density is for a gas disk of 1700AU and the
actual regions from where we expect the H$_{2}$ emission are well below a
radius of 200-300AU.  On the other hand, the UV flux scaling parameter
$\chi$ for an M0 type star is much less than 1. From these, the UV
fluorescence is expected to be of little significance for the excitation
of H$_{2}$ lines.

\begin{figure}
\resizebox{\hsize}{!}{\includegraphics{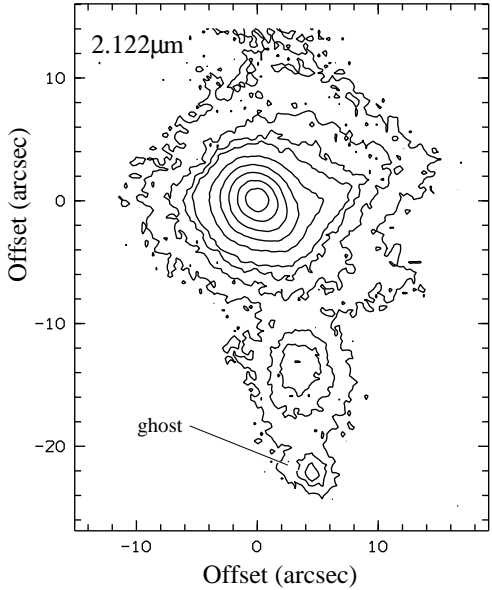}}

\caption[]{Contour plot of RNO91 taken through narrowband
 filter, illustrating an east-west disk associated with the source, as
 well as nebulous emission to the south. The contours measure
 5,10,15,20,40,80,160,320 and 640$\times$ the standard deviation to the
 mean background level in the image.}

\label{mag3} 
\end{figure}

Figure 3 shows a narrow band image through H$_{2}$ 
filter, without continuum subtraction. The disk structure and an outflow
lobe in the south can be seen clearly. These structures are more evident
in this narrow band image, than in the earlier K$^{\prime}$ image of
Hoddap (\cite{hod}) or the polarimetric image of W94. W94 had shown the
existence of frozen H$_{2}$O, CO and possibly XCN in a disk structure of
radius 1700AU around RNO\,91. The existence of these ices on grains within
this disk structure was proven by the absorption features found in a
$3-5\mu$m spectra. Their K-band polarimetric map also conforms with
models of scattering from disks (Whitney \& Hartmann \cite{whit}). W94
suggest a scenario with a flared disk viewed at an angle of $30\degr$ with
the northern outflow lobe tilted away from the observer. The narrow band
image shown in figure 3 (note specifically the patch of continuum
emission to the south of RNO\,91)  clearly support this scenario.  In
addition, it can be seen from fig 1 that the H$_{2}$ emission is extended
more in the southern direction than in the northern direction which also
confirms the tilt of the outflow axis, the northern lobe of the outflow being 
obscured near the source by the disk structure. 

\section{Discussion}
From the spectrum we measure an H$_{2}$ flux of 7$\times
10^{-14}$ ergs sec$^{-2}$ cm$^{-2}$ (integrated over 3 rows, representing
an area on-source of $2\arcsec \times 3\arcsec $).
If we attribute this flux to shocks, we can estimate the mass
flow rate $\dot{M}$ using the relation

   \begin{equation}
        \epsilon L = GM\dot{M}/ R  = 1/2\dot{M}v^{2} 
   \end{equation}
 
\noindent where $\epsilon$ is the ratio of the total energy in the shock
to the strength of the 1-0 S(1) line, $L$ is the observed H$_{2}$ line
luminosity, and $v$ the shock velocity.  We adopt a value for $v$ of 30 km
s$^{-1}$, a value that is optimum for producing H$_2$ line emission.  We
also assume that $\epsilon = 50$ (Smith \cite{smth}).  Together, these
yield an estimate for mass flux of $\dot{M} = 4\times 10^{-8} $ M$_{\odot}$
yr$^{-1}$. 

As demonstrated in the previous section, the relative change in the
intensity of H$_{2}$ from that of the photospheric features in the spectra
extracted with different widths of the continuum strip shows that   the
H$_2$ emission close to the star likely originates from a region 
different to that traced by the photospheric lines. The fact that the
H$_2$ line emission is extended along our N-S slit  strongly suggests that
the H$_2$ is associated with an outflow. By observing line emission
coincident with the RNO\,91 stellar continuum, we  may therefore be
tracing the outflow all the way back to the source. The total flux
measured from the observed extended emission (integrated over 9 rows
representing an area on-source of $2\arcsec \times 9\arcsec$) is
1.5$\times 10^{-13}$ ergs sec$^{-2}$ cm$^{-2}$. This represents an average
flux distribution of 3.5$\times 10^{-4} $ ergs sec$^{-2}$ cm$^{-2}
$sr$^{-2}$.  Assuming $L_{H2} \sim 10\times L_{S(1)}$ (Davis \&
Eisl\"offel \cite{dave}), the extinction corrected S(1) line flux yields a
net H$_{2}$ luminosity of 0.002L$_\odot$. Using the estimated value of
$\dot{M}$ and the kinematic age of the outflow, the H$_{2}$ luminosity can
be shown to represent a net warm H$_{2}$ mass of $\sim 5\times 10^{-4}$
M$_{\odot}$, where, a $9\arcsec$ long flow with a velocity of 30 km
s$^{-1}$ is used to calculate the age of the outflow.

The H$_2$ line flux measured from within the central 3$\arcsec$  could
be associated with the near-IR counterparts of  HH knots embedded within
the cocoon  that are excited by the outflow and accretion shocks in a
disk. However, excitation of all of the observed ``on-source'' 
H$_{2}$ emission due to infall is not feasible, since the mass accretion
rates derived from such infall will produce a K-band extinction (Herbst et
al.  \cite{her95}) that is an order of magnitude higher than the measured
value by Myers et al. (\cite{myr}). But some fraction of the
observed emission could originate from accretion shocks, with the rest
from the outflow, as in the case of T Tauri. It is interesting to note
that the high resolution images obtained via Shift and Add imaging at
UKIRT by A97 reveal a nebulous feature which is about $1\arcsec$ north of
the star. This could be a shock excited feature in the outflow. 
Alternatively, this feature -- in conjunction with the fainter features to
the S-W of the star -- appears to be part of an ellipse whose major axis
is $\sim$ 200AU as measured by us using the published images of A97. In
fact these features may be visualized  in terms of a disk tilted by
30$\degr$  to the north. The absence of any feature to the south may be 
attributed to the obscuration of the disk by the outflow. These arguments
support the W94 data and their  model, as do our narrow band images shown
in Figure 3.

While gas and dust disks are known to exist around PMS stars (Beckwith et
al. \cite{beck90}, Dutrey et al. \cite{dut}), there exists a small sample
of objects with disk structures that contain {\it icy mantles} where ices
are found on dust particles in the protostellar envelopes (see Chiar et
al. \cite{chiar} and references therein). RNO\,91 possesses by far the
largest icy gas/dust disk structure (20\arcsec) and also shows shocked
H$_{2}$ originating in an outflow and possibly an accretion disk. These
features make RNO\,91 unique and an interesting object for further studies
in millimeter and infrared wavelengths.

\section{Conclusions}

An infrared spectrum of RNO\,91 shows  emission of shocked molecular
hydrogen from an outflow in the N-S direction and from a spatially
unresolved region close to RNO\,91. We estimate a mass flow rate of
$\dot{M} = 4\times 10^{-8} $ M$_{\odot}$ yr$^{-1}$, based on the line
fluxes from this spatially unresolved region around RNO\,91. The line
fluxes from the spatially extended outflow yield $L_{H2} \sim
0.002$L$_{\odot}$ representing net warm H$_{2}$ mass of $\sim 5\times
10^{-4}$ M$_{\odot}$. The outflow seen here in H$_{2}$ emission,
extending roughly N-S, appears to support the tilted disk $+$ outflow
model of W94, where the northern flow lobe is tilted away from us at an
angle of $30\degr$ to the plane of the sky. Our narrow band images also
support this scenario. We suggest that  the H$_{2}$ flux from the
spatially unresolved region around the source could originate in
``HH-type'' knots embedded in the cocoon  surrounding the star. However,
accretion shocks cannot entirely be ruled out based on our observations.
We argue that RNO\,91 is a unique disk/outflow system.

\begin{acknowledgements} 
This work is supported by the Department of Space, Government of India.
The United Kingdom Infrared Telescope is operated by the Joint Astronomy
Centre on behalf of the U.K. Particle Physics and Astronomy Research
Council. The Imaging data reported here were obtained as part of the UKIRT
Service Programme (obtained for us by Antonio Chrysostomou). 
\end{acknowledgements}

\end{document}